\documentclass[showpacs,aps,graphicxt,twocolumn]{revtex4}
\usepackage{graphicx}

\begin{document}
\title{Efficient high-capacity quantum secret sharing with two-photon entanglement\footnote{Published in \emph{Phys. Lett.} A 372 (2008) 1957}}
\author{ Fu-Guo Deng,$^{1,2,3,4}$
Xi-Han Li,$^{1,2,3}$ and Hong-Yu Zhou$^{1,2,3}$}
\address{$^1$ The Key Laboratory of Beam Technology and Material
Modification of Ministry of Education, Beijing Normal University,
Beijing 100875,
People's Republic of China\\
$^2$ Institute of Low Energy Nuclear Physics, and Department of
Material Science and Engineering, Beijing Normal University,
Beijing 100875, People's Republic of China\\
$^3$ Beijing Radiation Center, Beijing 100875,  People's Republic of
China\\
$^4$ Department of Physics, Applied Optics Beijing Area Major
Laboratory, Beijing Normal University, Beijing 100875, People's
Republic of China}
\date{\today }

\begin{abstract}
An efficient high-capacity quantum secret sharing scheme is proposed
following some ideas in quantum dense coding with two-photon
entanglement. The message sender, Alice prepares and measures the
two-photon entangled states, and the two agents, Bob and Charlie
code their information on their photons with four local unitary
operations, which makes this scheme more convenient for the agents
than others. This scheme has a high intrinsic efficiency for qubits
and a high capacity.

\end{abstract}

 \maketitle

Quantum entanglement offers some novel ways for information
processing and transmitting securely \cite{book}, such as quantum
computation \cite{book}, quantum teleportation \cite{teleportation},
quantum key distribution (QKD)
\cite{Ekert91,BBM92,Gisin,LongLiu,CORE}, quantum dense coding
\cite{BW},  quantum secure direct communication (QSDC)
\cite{two-step,Wangc}, and so on. A surprising property of an
entangled quantum system is its nonlocality. Two parts of the
quantum system cannot be considered to be independent even if they
are far apart, and the single-particle measurements on these two
parts cannot give all the information about the state of the whole
quantum system. Quantum nonlocality has been embodied in the process
of quantum teleportation \cite{teleportation}, an important quantum
technique. Ekert exploited the nonlocality feature to design a QKD
protocol \cite{Ekert91} in 1991, and Bennett, Brassard and Mermin
(BBM92) \cite{BBM92} simplified its error rate analysis process in
1992. Also, quantum nonlocality has been used to transmit a secret
message directly \cite{two-step,Wangc}.

Secret sharing is a classical cryptographic scheme
\cite{Blakley,Blakley2,Blakley3} in which a boss, say Alice suspects
that one of her two remote agents, say Bob and Charlie, may be
dishonest but she does not know who the dishonest one is. She
believes that the honest agent can prevent the dishonest one from
destroying her benefits if they act in concert. For the security of
her message $M_A$, Alice splits it into two pieces $M_B$ and $M_C$,
and sends them to Bob and Charlie, respectively. The two agents can
read out the message $M_A=M_B\oplus M_C$ only when they cooperate.
As a classical signal can be copied perfectly, it is impossible to
create a private key with classical physics. When quantum mechanics
enters the field of information, the story is changed. Quantum
secret sharing (QSS) is the generalization of classical secret
sharing into quantum scenario and has progressed quickly in recent
years \cite{HBB99,KKI,Bandyopadhyay,Karimipour,guoqss,longqss,
Peng,deng2005,MZ,zhoup,chenp,dengmQSTS,dengpra,lixhQSTS,zhanglm,improving,cleve,YanGao,TZG,AMLance}.

One of the main goals of QSS, similar to QKD, is to distribute the
private keys among the three participants, or more generally, many
participants securely
\cite{HBB99,KKI,Bandyopadhyay,Karimipour,guoqss,deng2005,longqss,YanGao,MZ,zhoup,chenp,zhanglm,improving}.
An original QSS scheme was proposed by Hillery, Bu\v{z}ek and
Berthiaume \cite{HBB99} in 1999, which is called HBB99 hereafter. In
the HBB99 scheme \cite{HBB99}, the secret sharing is accomplished by
using a three-photon entangled Greenberger-Horne-Zeilinger (GHZ)
state. Each participant holds a photon from a GHZ state, and chooses
randomly one measuring-basis (MB) from the $X$-MB and the $Y$-MB to
measure their photons independently, similar to BBM92 QKD scheme
\cite{BBM92}. Subsequently Karlsson, Koashi and Imoto (KKI) put
forward another QSS scheme \cite{KKI} with a two-photon
polarization-entangled state. The photons are polarized along the
$z$ or $x$ directions, and the two agents measure their photons
choosing randomly one of the two MBs, $Z$ and $X$. The efficiency
for qubits $\eta_q\equiv \frac{q_u}{q_t}$ in these two protocols
\cite{HBB99,KKI} is 50\% because half of the instances are discarded
as the participants choose incompatible MBs, similar to the BBM92
QKD protocol \cite{BBM92}.  Here $q_u$ is the useful qubits and
$q_t$ is the total qubits transmitted \cite{LongLiu,Cabello}. Each
entangled quantum system can be used to carry on average a half bit
of the random common key, and two or more bits of classical
communications are required to compare the correlation of  their
MBs. Their total efficiency $\eta_t$ is low; $\eta$ is defined as
\cite{LongLiu,Cabello}
\begin{eqnarray}
\eta_t=\frac{b_s}{q_t+b_t},
\end{eqnarray}
where $b_s$, $q_t$ and $b_t$ are the number of bits in the raw key,
the qubits transmitted, and the total classical bits exchanged
between the participants in the quantum communication, respectively.
For instance, the total efficiency in HBB99 QSS scheme \cite{HBB99}
with three parties is at most $\eta_t=\frac{0.5}{2+2}=12.5\%$ as the
three parties transmit a two-qubit (i.e., two particles in a
three-particle GHZ state) quantum system to create half bit of key
at the expense of exchanging at least two bits of classical
information about their MBs. That is, Alice and Bob (or Charlie)
exchange a bit of information about Bob's (Charlie's) MB, i.e.,
$b_t=2$. In KKI QSS scheme \cite{KKI} with three parties, the
parties use a two-qubit quantum system to obtain half bit of key in
principle, i.e., its total efficiency is at most
$\eta_t=\frac{0.5}{2+2}=12.5\%$.

In this Letter, we present an efficient high-capacity QSS  scheme
for distributing a random key among three participants with a
two-photon entangled state based on quantum dense coding. The two
agents, Bob and Charlie choose the single-photon measurements on the
sampling photons with the three MBs $Z$, $X$ and $Y$ randomly for
eavesdropping check, and encode their information with four local
unitary operations on their photons. Almost all the entangled states
can be used to exchange the random key and each two-photon entangled
state can carry two bits of information. Moreover, this scheme is
secure with the decoy photons and the classical information
exchanged is reduced largely as the participants almost need not
compare their MBs for all the instances except for those for
eavesdropping check. The efficiency for qubits $\eta_q$  approaches
1 and the total efficiency $\eta_t$ approaches 50\% (neglecting the
instances for checking eavesdropping, same as those in other QSS
schemes) as the two qubits are transmitted double the distance
between the sender Alice and her agents, which equals four qubits
are transmitted in the KKI QSS scheme, and two bits of key are
created in theory.

An Einstein-Podolsky-Rosen (EPR) pair  is in one of the
four Bell states shown as follows.
\begin{eqnarray}
\vert \phi ^{+}\rangle &=&\frac{1}{\sqrt{2}}(\vert +z\rangle
_{B}\vert +z\rangle _{C} + \vert -z\rangle _{B}\vert -z\rangle
_{C})\nonumber\\
&=&\frac{1}{\sqrt{2}}(\vert +x\rangle _{B}\vert +x\rangle _{C} +
\vert -x\rangle _{B}\vert -x\rangle _{C})\nonumber\\
&=&\frac{1}{\sqrt{2}}(\vert +y\rangle _{B}\vert -y\rangle _{C} +
\vert -y\rangle _{B}\vert +y\rangle _{C}), \label{EPR4}
\nonumber\\
 \vert \phi
^{-}\rangle &=&\frac{1}{\sqrt{2}}(\vert +z\rangle _{B}\vert
+z\rangle _{C} - \vert -z\rangle _{B}\vert -z\rangle
_{C})\nonumber\\
&=&\frac{1}{\sqrt{2}}(\vert +x\rangle _{B}\vert -x\rangle _{C} +
\vert -x\rangle _{B}\vert +x\rangle _{C})\nonumber\\
&=&\frac{1}{\sqrt{2}}(\vert +y\rangle _{B}\vert +y\rangle _{C} +
\vert -y\rangle _{B}\vert -y\rangle _{C}), \label{EPR3}
\nonumber\\
\vert \psi ^{+}\rangle &=&\frac{1}{\sqrt{2}}(\vert +z\rangle
_{B}\vert -z\rangle _{C} + \vert -z\rangle _{B}\vert +z\rangle
_{C})\nonumber\\
&=&\frac{1}{\sqrt{2}}(\vert +x\rangle _{B}\vert +x\rangle _{C} -
\vert -x\rangle _{B}\vert -x\rangle _{C})\nonumber\\
&=&\frac{-i}{\sqrt{2}}(\vert +y\rangle _{B}\vert +y\rangle _{C} -
\vert -y\rangle _{B}\vert -y\rangle _{C}), \label{EPR2}
\nonumber\\
\vert \psi ^{-}\rangle &=&\frac{1}{\sqrt{2}}(\vert
+z\rangle _{B}\vert -z\rangle _{C} - \vert -z\rangle _{B}\vert
+z\rangle
_{C})\nonumber\\
&=&\frac{1}{\sqrt{2}}(\vert -x\rangle _{B}\vert +x\rangle _{C} -
\vert +x\rangle _{B}\vert -x\rangle _{C})\nonumber\\
&=&\frac{i}{\sqrt{2}}(\vert +y\rangle _{B}\vert -y\rangle _{C}
-\vert -y\rangle _{B}\vert +y\rangle _{C}), \label{EPR1}
\end{eqnarray}%
where $\vert +z\rangle \equiv \vert 0\rangle $ and $\vert -z\rangle
\equiv \vert 1\rangle $\ are the eigenvectors of the MB $Z$ (for
example the polarizations of a photon along the z-direction), and
$\vert +x\rangle \equiv \frac{1}{\sqrt{2}}(\vert 0\rangle + \vert
1\rangle)$ and $\vert -x\rangle \equiv \frac{1}{\sqrt{2}}(\vert
0\rangle - \vert 1\rangle)$\ are those of the MB $X$. The subscripts
$B$ and $C$ indicate the two correlated photons in each
Einstein-Podolsky-Rosen (EPR) pair. The four local unitary
operations $U_i$ $(i=0,1,2,3)$ can transform one of the Bell states
into another,
\begin{eqnarray}
U_{0}&&\equiv I=\vert 0\rangle \langle 0\vert + \vert 1\rangle
\langle 1\vert ,  \label{O0}
\nonumber\\
U_{1}&&\equiv \sigma _{z}=\vert 0\rangle \langle 0\vert -\vert
1\rangle \langle 1\vert ,  \label{O1}
\nonumber\\
U_{2}&&\equiv \sigma _{x}=\vert 1\rangle \langle 0\vert + \vert
0\rangle \langle 1\vert ,  \label{O2}
\nonumber\\
U_{3}&&\equiv i\sigma _{y}=\vert 0\rangle \langle 1\vert -\vert
1\rangle \langle 0\vert, \label{O3}
\end{eqnarray}%
where $I$ is the $2\times 2$ identity matrix and $\sigma_i$ $(i =
x, y, z)$ are the Pauli matrices, i.e.,
\begin{eqnarray}
&&I \otimes U _{0}\vert \psi^\pm\rangle=\vert \psi^\pm\rangle,
\;\;\;\;\;\; I \otimes U _{0}\vert \phi^\pm\rangle=\vert
\phi^\pm\rangle,\label{L0}\nonumber\\
&&I \otimes U _{1}\vert \psi^\pm\rangle=-\vert \psi^\mp\rangle,
\;\;\; I \otimes U _{1}\vert \phi^\pm\rangle=\vert
\phi^\mp\rangle,\label{L1}\nonumber\\
&&I \otimes U _{2}\vert \psi^\pm\rangle=\vert \phi^\pm\rangle,
\;\;\;\;\;\; I \otimes U _{2}\vert \phi^\pm\rangle=\vert
\psi^\pm\rangle,\label{L2}\nonumber\\
&&I \otimes U _{3}\vert \psi^\pm\rangle=\vert \phi^\mp\rangle,
\;\;\;\;\;\; I \otimes U _{3}\vert \phi^\pm\rangle=-\vert
\psi^\mp\rangle.\label{L3}
\end{eqnarray}

The four Bell states can be used to carry two bits of classical
information \cite{BW,two-step}, but we cannot send the photon pair
directly into an insecure channel as the four Bell states are the
the simultaneous eigenvectors of the two-body operators $\{ \sigma
_{z}^{(B)}\sigma _{z}^{(C)},\sigma _{x}^{(B)}\sigma _{x}^{(C)}\} $
and they can be copied freely, which renders the transmission
insecure. To prevent an eavesdropper Eve from eavesdropping, one way
is not allowing Eve to acquire simultaneously both photons in each
EPR pair, such as those in the two-step quantum communication scheme
\cite{LongLiu,two-step} and its variant \cite{Wangc}. Another method
is to change the order of a group of EPR pairs with two quantum
channels so as to confuse Eve the correct matching of the photons in
the group of EPR pairs, for instance that in the
controlled-order-rearrangement-encryption technique for QKD
\cite{CORE}.

In QSS, if the participants can prevent the dishonest agent, say
Bob, from eavesdropping the quantum channel freely,   any
eavesdropper can be found out \cite{KKI}. Similar to Ref.
\cite{KKI}, Alice can prepare each photon pair in one of the
nonorthogonal entangled states, which will forbid the dishonest one
to copy the quantum system without disturbing it. To the end, Alice
should pick out two nonorthogonal bases to prepare the entangled
photon pairs, similar to Bennet-Brassard 1984 (BB84) QKD protocol
\cite{BB84}. Certainly, one set of basis for an entangled photon
pair is the four Bell states, shown in Eq (\ref{EPR1}). Another set
of basis can be chosen as follows.
\begin{eqnarray}
\vert \Phi ^{+}\rangle &=&\frac{1}{\sqrt{2}}(\vert +x\rangle
_{B}\vert +z\rangle _{C} + i\vert -x\rangle _{B}\vert -z\rangle
_{C})
\nonumber\\
&=&\frac{1}{\sqrt{2}}(\vert +z\rangle _{B}\vert +y\rangle _{C} +
\vert -z\rangle _{B}\vert -y\rangle
_{C})\nonumber\\
&=&\frac{e^{\frac{-i\pi}{4}}}{\sqrt{2}}(\vert +y\rangle _{B}\vert
-x\rangle _{C} + i\vert -y\rangle _{B}\vert +x\rangle
_{C}), \label{EPR11} \nonumber\\
\vert \Phi ^{-}\rangle &=&\frac{1}{\sqrt{2}}(\vert +x\rangle
_{B}\vert +z\rangle _{C} -i\vert -x\rangle _{B}\vert -z\rangle _{C})
\nonumber\\
&=& \frac{1}{\sqrt{2}}(\vert +z\rangle _{B}\vert -y\rangle _{C} +
\vert -z\rangle _{B}\vert +y\rangle
_{C})\nonumber\\
&=&\frac{e^{\frac{-i\pi}{4}}}{\sqrt{2}}(\vert +y\rangle _{B}\vert
+x\rangle _{C} + i\vert -y\rangle _{B}\vert -x\rangle _{C}),
\label{EPR13}
\nonumber\\
\vert \Psi ^{+}\rangle &=& \frac{1}{\sqrt{2}}(\vert +x\rangle
_{B}\vert -z\rangle _{C} + i\vert -x\rangle _{B}\vert +z\rangle
_{C})
\nonumber\\
&=& \frac{i}{\sqrt{2}}(\vert +z\rangle _{B}\vert -y\rangle _{C} -
\vert -z\rangle _{B}\vert +y\rangle
_{C})
\nonumber\\
&=&\frac{e^{\frac{i3\pi}{4}}}{\sqrt{2}}(\vert +y\rangle _{B}\vert
-x\rangle _{C} - i\vert -y\rangle _{B}\vert +x\rangle _{C}), \label{EPR12}\nonumber\\
\vert \Psi ^{-}\rangle &=&\frac{1}{\sqrt{2}}(\vert +x\rangle
_{B}\vert -z\rangle _{C} - i\vert -x\rangle _{B}\vert +z\rangle
_{C})\nonumber\\
&=& \frac{-i}{\sqrt{2}}(\vert +z\rangle _{B}\vert +y\rangle _{C} -
\vert -z\rangle _{B}\vert -y\rangle _{C})\nonumber\\
&=&\frac{e^{\frac{-i\pi}{4}}}{\sqrt{2}}(\vert +y\rangle _{B}\vert
-x\rangle _{C} + i\vert -y\rangle _{B}\vert +x\rangle _{C}),
\label{EPR14}
\end{eqnarray}
where $\vert \pm y\rangle=\frac{1}{\sqrt{2}}(\vert 0\rangle \pm
i\vert 1\rangle)$ are the two eigenvectors of the MB Y. The four
local unitary operations $U_i$ $(i=0,1,2,3)$ can transfer one of the
four entangled states $\{\vert \Phi^\pm\rangle, \vert
\Psi^\pm\rangle\}$ into another, i.e.,
\begin{eqnarray}
&&I \otimes U _{1}\vert \Phi^\pm\rangle=\vert \Phi^\mp\rangle,
\,\,\,\,\,\,\,\,\, I \otimes U _{1}\vert \Psi^\pm\rangle=\vert
\Psi^\mp\rangle,\label{L20}
\nonumber\\
&&I \otimes U _{2}\vert \Phi^\pm\rangle= \vert \Psi^\pm\rangle,
\,\,\,\,\,\,\,\,\, I \otimes U _{2}\vert \Psi^\pm\rangle= \vert
\Phi^\pm\rangle,\label{L21}
\nonumber\\
&&I \otimes U _{3}\vert \Phi^\pm\rangle=-\vert \Psi^\mp\rangle,
\,\,\,\,\, I \otimes U _{3}\vert \Psi^\pm\rangle=\vert
\Phi^\mp\rangle, \label{L22}
\nonumber\\
 &&U _{1} \otimes I\vert
\Phi^\pm\rangle=\vert \Psi^\mp\rangle, \,\,\,\,\,\,\,\,\, U _{1}
\otimes I\vert \Psi^\pm\rangle=\vert \Phi^\mp\rangle,\label{L30}
\nonumber\\
&&U _{2} \otimes I\vert \Phi^\pm\rangle=\vert \Phi^\mp\rangle,
\,\,\,\,\,\,\,\, U _{2} \otimes I\vert \Psi^\pm\rangle=-\vert
\Psi^\mp\rangle,\label{L31}
\nonumber\\
&&U _{3} \otimes I\vert \Phi^\pm\rangle=\vert \Psi^\pm\rangle,
\,\,\,\,\,\,\,\, U _{3} \otimes I\vert \Psi^\pm\rangle=-\vert
\Phi^\pm\rangle. \label{L32}
\end{eqnarray}
The two basis sets $\{\vert \phi^\pm\rangle, \vert
\psi^\pm\rangle\}$ and $\{\vert \Phi^\pm\rangle, \vert
\Psi^\pm\rangle\}$ are not orthogonal, which forbids any one to
copy them perfectly, same as that in Ref. \cite{KKI}.

Now, let us describe the principle of our QSS scheme in detail as
follows.

(1). Alice, Bob and Charlie agree that each of the four local
unitary operations $U_0$, $U_1$, $U_2$ and $U_3$ represents the
two-bit information.

(2). Alice prepares the two photons $B$ and $C$ in one of the
eight nonorthogonal entangled states $\{\vert \phi^\pm\rangle$, $
\vert \psi^\pm\rangle$, $\vert \Phi^\pm\rangle$, $\vert
\Psi^\pm\rangle\}$ randomly. She sends the photon $B$ to Bob and
$C$ to Charlie.

For preventing the dishonest agent from eavesdropping freely with an
opaque attack \cite{fakesignal}, Alice sends a decoy photon
\cite{decoy-photon,lixhjkps}, which is randomly in one of the six
states $\{\vert 0\rangle, \vert 1\rangle, \vert +x\rangle, \vert
-x\rangle, \vert +y\rangle, \vert -y\rangle\}$, to each agent with
the probability $p_d$ (we give the reason for choosing decoy photons
below). Alice can prepare the decoy photon by measuring one photon
in the two-photon quantum system which is in one of the two states
$\{\vert \phi^+\rangle, \vert \Phi^+\rangle\}$ with the MB
$\sigma_z$ \cite{lixhjkps}. Also, she can produce it with an ideal
single-photon source \cite{decoy-photon,lixhjkps}.

(3). Bob and Charlie choose one of the two modes, a small
probability $p_c $ ($<1/2$) with the checking-eavesdropping mode and
a large probability $1-p_c$ with the coding mode, for their photons
received, similar to those in the Refs. \cite{two-step,Wangc,QOTP}.

If Bob (Charlie) chooses the checking-eavesdropping mode, Bob
(Charlie) measures his  photon by choosing one of the three MBs $Z$,
$X$ and $Y$ randomly; otherwise, Bob (Charlie) encodes his random
key on the photon received with one of the four unitary operations
$\{U_i\}$ $(i=0,1,2,3)$. He sends the photon back to the sender
Alice.

(4) Alice takes a Bell-basis measurement on each two correlated
photons received from Bob and Charlie with the two-photon
entanglement basis $\{\vert\phi^\pm\rangle, \vert\psi^\pm\rangle\}$
or $\{\vert\Phi^\pm\rangle, \vert\Psi^\pm\rangle\}$, as  the same as
that she prepares them before the communication.

As the operations done by the agents on the quantum system composed
of the photons $B$ and $C$ do not change its basis, the measurement
done by Alice is deterministic and will give out the outcome of the
combination of the unitary operations performed by Bob and Charlie,
say $U_A = U_B \otimes U_C$. Here $U_B$ and $U_C$ are the operations
done by Bob and Charlie, respectively. If the communication is
secure, each of the four unitary operations represents two bits of
classical information which can be used as the raw key in QSS.

If one of the two agents, say Bob, measures his photon and the other
agent (Charlie) sends his photon back to Alice, instead of encoding
it, Alice will get nothing with her Bell-basis measurement.

(5). Alice tells her agents which entangled basis has been chosen
for each EPR pair and completes the error rate with the helps of
her two agents.

She requires Bob and Charlie to publish the MBs and the outcomes of
the sample photons for which they choose the checking-eavesdropping
mode. Alice exploits the refined error analysis technique \cite{ABC}
for checking eavesdropping of the process of the transmission from
Alice to her agents. That is, Alice only picks up the decoy photons
measured by the agents to check eavesdropping. As the agents measure
the decoy photons with the three MBs, $Z$, $Y$ and $X$, the
probability that the outcomes of the agents' are correlated with
those of Alice's is $\frac{1}{3}P_dP_c$. Similar to Ref. \cite{ABC},
this eavesdropping check will find out the eavesdropper monitoring
the quantum channel from Alice to her agents as any eavesdropping
will leave a trace in the outcomes of the decoy sampling photons.

For preventing the dishonest agent from eavesdropping the process
from the other agent to Alice freely, Alice should also pick out
randomly a sufficiently large subset of the outcomes from the
Bell-basis measurements on the entangled quantum systems, and
analyzes its error rate, named it as the second check. It is useful
for check the security of the quantum channel when the photons run
from the two agents back to Alice. Moreover, it can provide an
estimate information for the postprocessing, such as the error
correct and the privacy amplification. For half of these instances,
Alice requires Bob first publish his operations and then Charlie, or
vice versa.

(6). If all the error rates are low than the threshold
$\varepsilon_{th}$, they can use the results remained  as a raw key
and distill a private key $K_A=K_B \oplus K_C$ with error correction
and privacy amplification \cite{Gisin}; otherwise, they will abandon
the outcomes transmitted and repeat the quantum communication from
the beginning.

It is of interesting to point out the advantages that Alice replaces
some photons in the Bell states with the decoy photons
\cite{decoy-photon,lixhjkps}. If Alice does not exploit the decoy
photons, a dishonest agent, say Bob can steal some information with
an opaque attack freely and fully \cite{fakesignal}, especially when
the transmission efficiencies lower than 50\%. In detail, Bob
intercepts the photon $C$ when it is sent from Alice to Charlie, and
stores it with a quantum memory. He prepares a fake EPR pair $B'C'$
in the state $\vert \phi^+\rangle_{B'C'}=\frac{1}{\sqrt{2}}(\vert
+z\rangle_{B'} \vert +z\rangle_{C'} + \vert -z\rangle_{B'}\vert
-z\rangle_{C'})$ and sends the photon $C'$ to Charlie, instead of
the photon $C$. If Charlie operates the photon $C'$ and sends it
back to Alice, Bob can capture this photon and take a Bell-state
measurement on the photons $B'C'$. He can obtain all the information
about the operations done by Charlie in this way. On the other hand,
if Charlie measures the photon $C'$ with the MB $Z$, $X$ or $Y$, Bob
will get no photon in the quantum signal sent from Charlie to Alice.
He can determine this condition with a quantum non-demolition
measurement, as same as that in Ref. \cite{two-step,QOTP}.
Subsequently, Bob can keep the photon $B$ only when he gets the
outcome $\vert \phi^+\rangle_{B'C}$. In this way, Bob's
eavesdropping introduces no errors in the outcomes of the
measurements done by Bob and Charlie (or Alice and Charlie) no
matter what the MB chosen by Charlie is \cite{fakesignal}.
Certainly, Bob's eavesdropping will introduce errors  in the
outcomes if Bob gets the other three Bell-basis results. However,
Bob can hide his eavesdropping with cheating \cite{fakesignal}. That
is, he can announce that he gets nothing when he measures the photon
$B$ as there are losses in the quantum line \cite{fakesignal}. On
the other hand,  without the decoy photons, Alice should measure one
EPR particle when the other EPR particle is measured by her agent.
For accomplishing this task, she should first judge the photon
number in each signal sent back by each agent, which will require
Alice to have the capability of taking a quantum non-demolition
measurement.

The process of eavesdropping check with decoy photons between Alice
and Charlie does not require Bob to participate in it, which will
forbid Bob eavesdrop the quantum channel from Alice to Charlie with
an opaque attack strategy \cite{fakesignal}. The same process takes
place between Alice and Bob. In essence, the process of the photon
transmission from Alice to her each agent is similar to the QKD
protocol proposed by Lo et al. \cite{ABC} which is proven to be
secure. That is, this process can be made to be secure, same as QKD
\cite{BB84,ABC}.

After the operations done by the agents, the density matrix of the
quantum system composed of the photons $B$ and $C$ is
$\rho_{BC}=\left(
\begin{array}{cccc}
1/4 & 0 & 0 & 0 \\
0 & 1/4 & 0 & 0 \\
0 & 0 & 1/4 & 0 \\
0 & 0 & 0 & 1/4%
\end{array}%
\right) $ for the dishonest agent, and he cannot copy the quantum
signal freely, similar to BB84 \cite{BB84} and BBM92 \cite{BBM92}
QKD protocols. So this QSS scheme can be made to be secure, which is
in principle different from the QSS schemes
\cite{HBB99,KKI,Bandyopadhyay,Karimipour,deng2005,longqss,zhanglm,improving,YanGao}
in which the process of checking eavesdropping is completed with the
cooperation of the dishonest agent. On the other hand, as the
security of the process of the transmission from Alice to her agents
is ensured with the decoy photons and that from the agents to Alice
is ensured with the samples of the outcomes obtained with Bell-basis
measurements, this QSS scheme does not require the participants to
have the capability of storing quantum states. As for the
multi-photon attack \cite{improving}, the agents can use a filter to
prevent a fake photon with a nonstandard wavelength from entering
their devices and use some beam splitters to split the sampling
signals chosen for eavesdropping check before they measure the
signals with the MB $Z$, $X$ or $Y$, same as that in Ref.
\cite{lixhpra}. Also, the parties can complete a faithful qubit
transmission against collective noise with the technique in Ref.
\cite{lixhapl}, which will improve the practical efficiency in this
QSS scheme.

Without the decoy photons prepared by Alice, this QSS scheme can be
made secure if both  Bob and Charlie have an ideal single-photon
source. In detail, when Bob (Charlie) chooses the
checking-eavesdropping mode, he   measures the photon $B$ ($C$) sent
by Alice with one of the three bases $Z$, $X$, and $Y$, and then
sends a photon prepared by himself  to Alice. This photon can be
randomly in one of the six states $\{\vert \pm z\rangle, \vert \pm
x\rangle, \vert \pm y\rangle\}$. In this way, the eavesdropper does
not know which is the sample photon before Alice receives the
entangled EPR pair, and the opaque attack can be overcome as Alice
can use the photons inserted by her agents to check the security of
the transmission of photons from the agents to Alice. In essence,
the honest agent prepares the decoy photons and uses them to prevent
the potentially dishonest agent from eavesdropping freely.

In fact, this QSS scheme is the modified version of the KKI QSS
scheme \cite{KKI} with quantum dense coding and decoy photons. But
only those modifications increase its intrinsic efficiency, the
source capacity  and the security largely. Almost all the instances
($(1-P_d)(1-P_c)^2$) are useful for generating the raw key except
for those chosen for eavesdropping check, and each of the two-photon
entangled quantum system can carry two bits of information.
Moreover, the classical information exchanged is reduced largely as
the two agents need not publish their MBs when they choose the
coding mode with the four local unitary operations. Then the
efficiency for qubits $\eta_q=(1-P_d)(1-P_c)^2$ approaches 1 when
$P_d$ and $P_c$ are very small. As the two qubits in the entangled
states are transmitted double the distances between the sender Alice
and her agents, which equals to that Alice and her agents transmit
four qubits, the total efficiency
$\eta_t=\frac{2}{4}\frac{(1-P_d)(1-P_c)^2}{1+P_c}$ in the present
QSS scheme, approaching 50\% when $P_d$ and $P_c$ are very small.
Certainly, in a practical application with a noisy quantum channel,
the efficiency for qubits  cannot approach 1 and the total
efficiency is low than 50\% as the probabilities $P_d$ and $P_c$
cannot be arbitrarily small. The parties can exploit error-avoiding
codes to reduce the effect of noise on the efficiencies, such as the
faithful qubit transmission technique with linear optics
\cite{lixhapl}.

In summary, we introduce an efficient high-capacity QSS scheme with
quantum dense coding based on two-photon entangled states. The two
agents, Bob and Charlie choose the single-photon measurements on the
sampling photons with the three MBs randomly for eavesdropping
check, and encode their information with four local unitary
operations, which make this QSS scheme more convenient for the
agents than some others \cite{HBB99,KKI}. Almost all the entangled
photon pairs can be used to exchange the random key and each photon
pair can carry two bit of information. The intrinsic efficiency for
qubit is double as that in KKI QSS scheme \cite{KKI}, and the source
capacity is four times as the latter with the photons running forth
and back. Moreover, this scheme is secure with decoy photons and the
classical information exchanged is reduced largely as the
participants almost need not compare their MBs for all the instances
except for those for eavesdropping check. As the efficiency for
producing three-particle entangled state is low than that for
two-particle entangled state, the present QSS scheme is more
practical with present technology than the first one proposed by
Hillery, Bu$\breve{z}$ek, and Berthiaume \cite{HBB99}.

\end{document}